\documentclass[a4paper,11pt]{article}
\usepackage{pos}
\renewcommand{\logo}{\relax}
\graphicspath{{./}{../}}

\newcommand{\order}{{\cal O}}
\newcommand{\One}{1\kern-4.5pt1}

\newcommand{\om}{\omega}

\newcommand{\lam}{\lambda}

\newcommand{\smkernel}[1]{\Bar\Delta(\omega,#1)} % smearing

\title{Spectral properties of bottomonium at high temperature: a systematic investigation}
\ShortTitle{Spectral beauty methods}

\author*[a,b]{Jon-Ivar Skullerud}
\author[c]{Gert Aarts}
\author[c]{Chris Allton}
\author[c]{M. Naeem Anwar}
\author[b]{Ryan Bignell}
\author[c]{Timothy J. Burns}
\author[a]{Rachel Horohan D'arcy}
\author[d]{Benjamin J\"ager}
\author[e]{Seyong Kim}
\author[f]{Maria Paola Lombardo}
\author[c]{Ben Page}
\author[b]{Sin\'ead M. Ryan}
\author[c]{Antonio Smecca}
\author[c]{Tom Spriggs}

\affiliation[a]{Department of Physics,
  Maynooth University---National University of Ireland Maynooth,
  Maynooth, Co Kildare, Ireland}
\affiliation[b]{School of Mathematics, Trinity College, Dublin, Ireland}
\affiliation[c]{Department of Physics, Swansea University,
       Singleton Park, Swansea SA2 8PP, UK
}
\affiliation[d]{
  Quantum Field Theory Center \& Danish IAS,
  Department of Mathematics and Computer Science
  University of Southern Denmark, 5230 Odense M, Denmark
  }
\affiliation[e]{
  Department of Physics, Sejong University, Seoul 05006, Korea
  }
\affiliation[f]{
INFN, Sezione di Firenze, 50019 Sesto Fiorentino (FI), Italy
}
\emailAdd{jonivar.skullerud@mu.ie}

\abstract{ We investigate spectral features of bottomonium at high
  temperature, in particular the thermal mass shift and width of
  ground state S-wave and P-wave state.  We employ and compare a range
  of methods for determining these features from lattice NRQCD
  correlators, including direct correlator analyses (multi-exponential
  fits and moments of spectral functions), linear methods
  (Backus-Gilbert, Tikhonov and HLT methods), and Bayesian methods for
  spectral function reconstruction (MEM and BR).  We comment on the
  reliability and limitations of the various methods.  }

\FullConference{The XVIth Quark Confinement and the Hadron Spectrum Conference (QCHSC24)\\
 19-24 August, 2024\\
 Cairns Convention Centre, Cairns, Queensland, Australia\\}

\begin{document}
\maketitle

\section{Introduction}

An important task in QCD phenomenology is to determine the spectral
functions of hadronic and current--current correlators.  These encode
physical information about \emph{inter alia} the existence and
properties of bound states and resonances, transport properties, and
multiparticle thresholds.  The spectral functions are related to the
euclidean correlators computable on the lattice by an integral
transform, but computing the spectral function $\rho(\om)$ given a
finite set of noisy correlator data $G(\tau)$ is known to be an
ill-posed problem.  Numerous methods have been developed to attempt to
handle this problem (see \cite{Rothkopf:2022fyo,Jay:2025ftw} for
recent reviews), which all have their inherent limitations.  The
purpose of the study that is reported here is to obtain a better
understanding of a subset of these methods by applying them to a
specific problem, namely determining the temperature dependence of the
mass and width of the 1S and 1P bottomonium states computed in lattice
non-relativistic QCD (NRQCD).

The bottomonium system has been studied in lattice NRQCD for a long
time
%\cite{Aarts:2010ek,Aarts:2011sm,Aarts:2014cda,Kim:2014iga,Kim:2018yhk,Larsen:2019bwy,Larsen:2019zqv,Larsen:2020rjk,Ding:2025fvo}.
\cite{Aarts:2011sm,Aarts:2014cda,Kim:2018yhk,Larsen:2019bwy,Larsen:2019zqv,Ding:2025fvo}.
The NRQCD approach has the benefit of a simple spectral representation with a
temperature-independent kernel,
\begin{equation}
  G(\tau) = \frac{1}{2\pi}\int_{\om_{\min}}^{\infty}e^{-\om\tau}\rho(\om)d\om\,.
  \label{eq:spectral}
\end{equation}
Since the heavy quark is not in thermal equilibrium, the
full range of points $\tau/a_\tau=0,\dots,N_\tau-1$ are available for
the spectral reconstruction.  Finally, since at least the
$\Upsilon(1S)$ state is expected to survive as a bound
state up to quite high temperatures, the peak position and thermal
width remain well-defined quantities which can be used to benchmark
the methods.

Earlier results from this project were presented in \cite{Spriggs:2021dsb}.
Some of the results shown here have previously been presented in
\cite{Smecca:2025hfw,Darcy:2025tzj,Bignell:2025bga}.

\section{Lattice setup}
\label{sec:lattice}

Our simulations are carried out using anisotropic lattices with an
$\order(a^2)$ improved gauge action and an $\order(a)$ improved Wilson
fermion action with stout smearing, following the parameter tuning and ensembles generated by the Hadron Spectrum Collaboration \cite{Edwards:2008ja,Lin:2008pr}.  The results presented here were
produced using the ``Gen2L'' ensembles, which have $N_f=2+1$ active
quark flavours with $m_\pi\approx240\,$MeV and an approximately
physical strange quark.  The spatial lattice spacing is
$a_s=0.112\,$fm and the anisotropy $\xi=a_s/a_\tau=3.45$.  The
temperature is given by $T=(a_\tau N_\tau)^{-1}$ and is varied by
changing the number of sites $N_\tau$ in the temporal direction.  For
more details about the ensembles, see Ref.~\cite{Aarts:2020vyb,Aarts:2022krz} and references therein.

\begin{table}[ht]
\begin{tabular}{|l|rrrrrrrrrrrr|}\hline
  $N_\tau$ & 
  128 &  64 &  56 &  48 &  40 &  36 &  32 &  28 &
  24 &  20 & 16 &  12  \\ \hline
  $T$ (MeV) & 47 & 95 & 109  & 127  & 152  & 169  & 190  & 217  &
  253  & 304  & 380  & 507  \\ %\hline
$T/T_{pc}$  & 0.28 & 0.57 & 0.65 & 0.76 & 0.91 & 1.01 & 1.14 & 1.30
 & 1.52 & 1.82 & 2.28 & 3.03 \\ \hline
\end{tabular}
\caption{Temporal extent $N_\tau$ and temperatures $T$ in MeV and in
  units of the pseudocritical temperature $T_{pc}$ for the Gen2L
  ensembles, used in this study.  Approximately 1000 configurations,
  separated by 10 HMC trajectories, were used at each temperature.}
\end{table}

The $b$ quarks were simulated using an NRQCD action including
$\order(v^4)$ and leading spin-dependent corrections, with mean-field
improved tree-level coefficients \cite{Aarts:2014cda}.

\section{Time-derivative moments}
\label{sec:moments}

The idea behind the moments method is to consider the correlator in
\eqref{eq:spectral} as the integral of a weighted spectral function
$\rho^\tau(\om)=e^{-\om\tau}\rho(\om)$.  The temporal derivatives of the
correlator are then the moments of $\rho^\tau(\om)$, and the salient
features of $\rho^\tau(\om)$, such as the locations and widths of its
peaks, and hence those of $\rho(\om)$, can be inferred from these
derivatives.

If $\rho(\om)$ is approximated by a sum of Gaussians,
\begin{align}
    \rho(\omega) &= \sum_{i=0}^{\infty}
    A_ie^{-\frac{(\omega-m_i)^2}{2\Gamma_i^2}},
    \intertext{we find}
 G(\tau) &= \sum_{i=0}^{\infty} A_i e^{-m_i\tau+\Gamma_i^2\tau^2/2} 
  = A_0e^{-m_0\tau+\Gamma_0^2\tau^2/2}\left(1+\sum_{i=1}^{\infty}\frac{A_i}{A_0}e^{-\Delta m_i\tau+\Delta\Gamma_i^2\tau^2/2}\right) \\
 \frac{d\log(G(\tau))}{d\tau}
 &\approx (-m_0+\Gamma^2_0\tau)
 - \sum_{i=1}^{\infty} \frac{A_i}{A_0}
 \Big(\Delta m_i-\Delta\Gamma_i\tau\Big)
 e^{-\Delta m_i \tau + \Delta\Gamma_i^2\tau^2/2} \label{eq:moments-mass}\\
\frac{d^2\log(G(\tau))}{d\tau}
&\approx \Gamma^2_0 + \sum_{i=1}^{\infty} B_i(\tau)e^{-\Delta m_i \tau
  + \Delta\Gamma_i^2\tau^2/2}\,,
\label{eq:moments-width}
\end{align}
where $B_i(\tau)$ can be approximated by a constant if
$\Delta\Gamma_i^2\tau\ll\Delta m_i$.  Specifically, from
\eqref{eq:moments-width} we see that the ground state width can be
extracted in the large-$\tau$ limit assuming the same condition holds,
while the first derivative \eqref{eq:moments-mass} can be used to
determine the ground state mass.

\begin{figure}[t]
\includegraphics[width=0.5\textwidth]{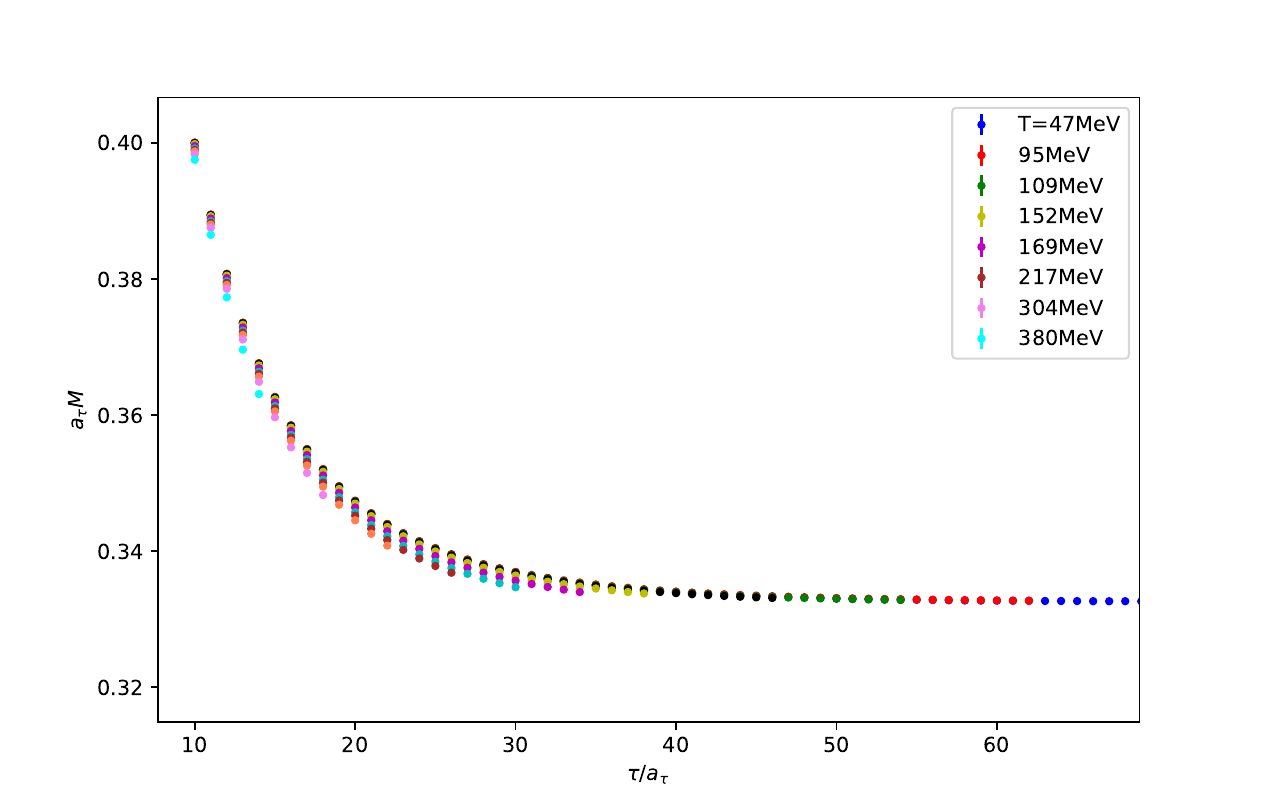}
\includegraphics[width=0.5\textwidth]{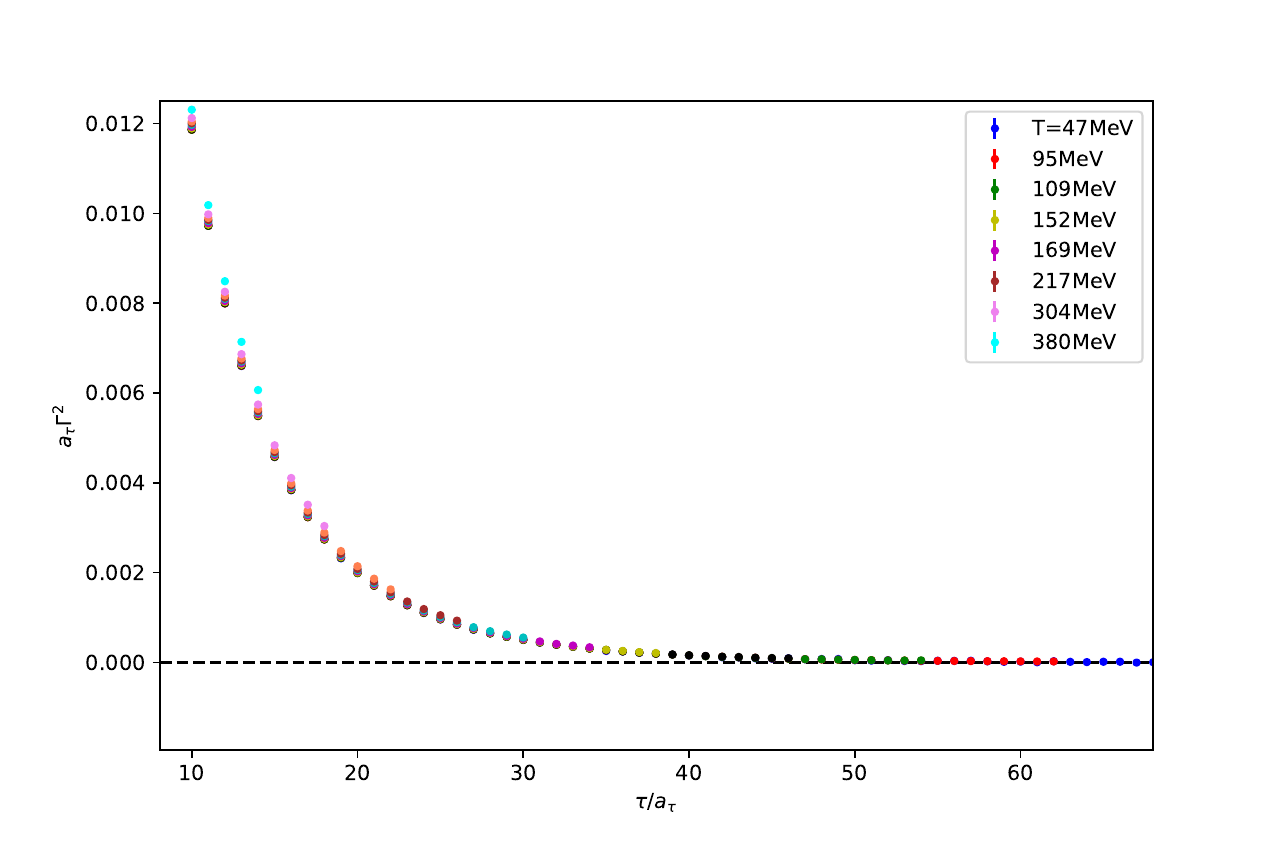}
\caption{First (left) and second (right) logarithmic derivative of the
  vector ($\Upsilon$) correlator at different temperatures.  Both
  quantities can be seen to approach a constant at large $\tau$,
  consistent with the expectations from
  eqs~(\ref{eq:moments-mass},\ref{eq:moments-width}), yielding the
  ground state mass and width respectively.  Note that since the
  second derivative is very close to zero, the linear decrease in the
  first derivative is not visible in the left-hand plot.}
\label{fig:moments-tau}
\end{figure}

The results for the vector ($\Upsilon$) channel
are shown in fig.~\ref{fig:moments-tau}.  We see that both the
first and second log-derivative approach a plateau at large $\tau$.
To determine the mass and width, the log-derivatives are fitted to a
simplified version of (\ref{eq:moments-mass},\ref{eq:moments-width}),
\begin{align}
  M(\tau) &= m_0 + Ae^{-c\tau} \,,&
  \Gamma^2(\tau) &= \Gamma_0^2 + Be^{-b\tau+d\tau^2}\,.
\end{align}

\begin{figure}[t]
\includegraphics[width=0.5\textwidth]{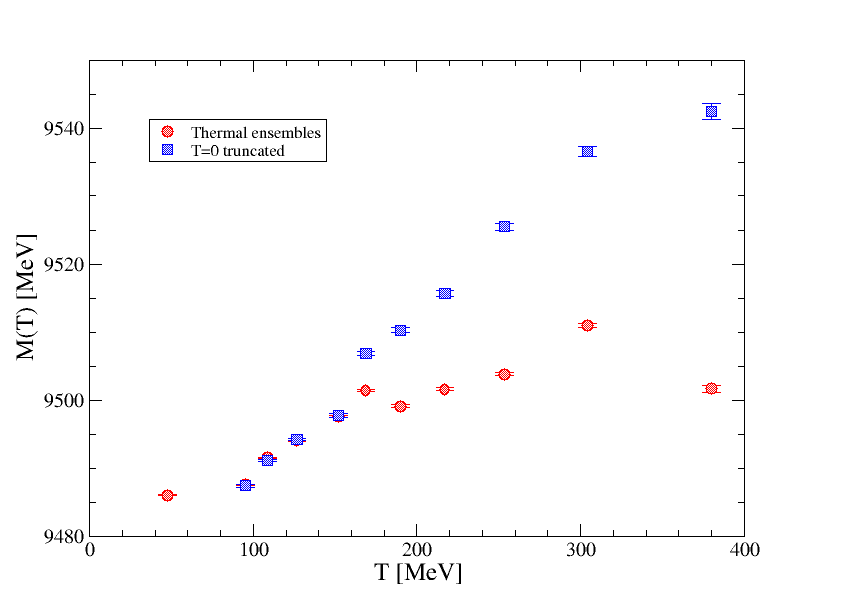}
\includegraphics[width=0.5\textwidth]{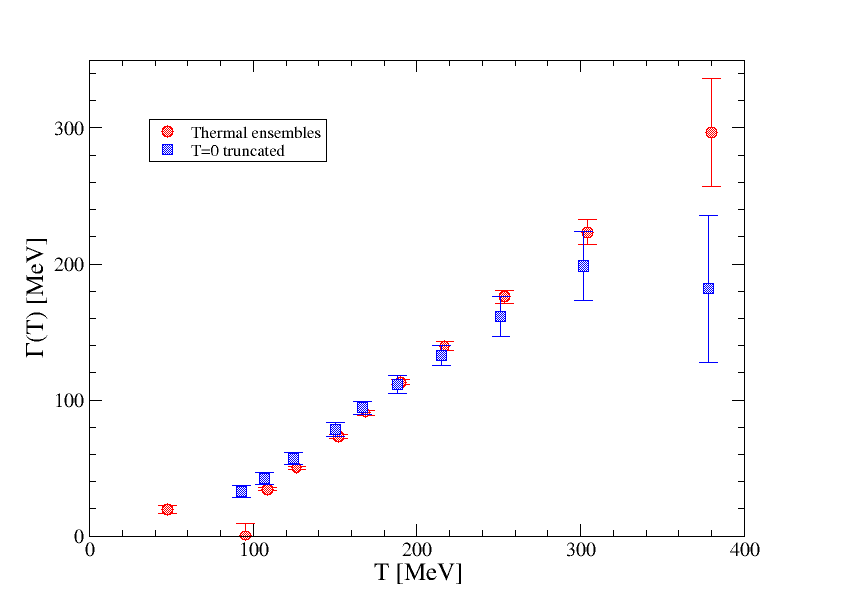}
\caption{Mass (left) and width (right) of the $\Upsilon$(1S) from the
  moments method, as functions of temperature $T$.  The red points
  show the results from the thermal ensembles, while the blue points
  show the results from the zero-temperature ensemble with the same
  temporal range (see text for details).}
\label{fig:moments-T}
\end{figure}
The results for $m_0$ and $\Gamma_0$ for $\Upsilon$(1S) are shown in
fig.~\ref{fig:moments-T} as a function of temperature (red points).  At first
sight, both $m$ and $\Gamma$ appear to increase with temperature.
However, this increase may in part be an effect of the shorter
temporal range available at higher temperatures, where $M(\tau)$ and
$\Gamma^2(\tau)$ have not yet reached a plateau at the maximum value
of $\tau$.  To disentangle this from real, thermal effects, the same
analysis has been carried out on the $N_\tau=128$ ensemble (which may
be considered to be at $T=0$), with the temporal range restricted to
$\tau_{\max}=1/T$.  The results of this are shown by the blue points
in fig.~\ref{fig:moments-T}.  We see that there are no substantial
thermal effects for $T\lesssim T_{pc}\approx170\,$MeV.  Above this
temperature, the mass from the thermal ensemble lies below that from
the corresponding truncated $T=0$ analysis, while the thermal width appears
to be increasing from zero at $T\gtrsim200\,$MeV (albeit still
consisten with zero within errors).  For further details about the method and
results, see \cite{Darcy:2025tzj}.

\section{Linear methods}
\label{sec:linear}

All the linear methods considered here (Tikhonov
\cite{tikhonovStabilityInverseProblems1943}, Backus--Gilbert 
\cite{backusResolvingPowerGross1968}, and HLT \cite{Hansen:2019idp})
regularise the inverse problem by avoiding a pointwise estimate of
$\rho(\om)$ and instead constructing a ``smeared'' spectral
function,
\begin{equation}
    \hat{\rho}(\omega_0) =\int_{\omega_\mathrm{min}}^{\omega_\mathrm{max}}\smkernel{\omega_0}\rho(\omega)d\omega,
    \label{eq:smeared-spectral}
\end{equation}
where the smearing kernel $\smkernel{\omega_0}$ plays the role of a
finite-width approximation to $\delta(\om-\om_0)$.  It may in turn 
be expressed as a linear combination of the NRQCD kernel functions,
\begin{align}
    \smkernel{\omega_0} &= \sum_\tau b_\tau(\omega_0)e^{-\omega\tau},\label{eq:backus_smkernel}
\end{align}
where $b_\tau(\omega_0)$ are coefficients which encode the location of the sampling point $\omega_0$ and are determined by the method. 
Inserting \eqref{eq:backus_smkernel} into
\eqref{eq:smeared-spectral} and using the spectral definition of the correlator \eqref{eq:spectral} then yields
\begin{equation}
    \hat{\rho}(\omega_0) = \sum_\tau b_\tau(\omega_0) G(\tau)\,.
    \label{eq:backus_spec_est_coeffs}
\end{equation}
The coefficients $b_\tau$ are determined by minimising a functional
$W[g_\tau]=A[g_\tau]+\lam B[g_\tau]$, where the three methods differ
by their choices for these functionals:
\begin{align}
  &\text{Tikhonov:}
&A[g_\tau] &= \int_{\om_{\min}}^{\infty}d\om 
    (\om-\om_0)^2[\Bar{\Delta}(\om,\om_0)]^2\,,
&B[g_\tau] &= \sum_{\tau_1,\tau_2}g_{\tau_1}g_{\tau_2}I(\tau_1,\tau_2)\\
&\text{BG:}
&A[g_\tau] &= \int_{\om_{\min}}^{\infty}d\om 
    (\om-\om_0)^2[\Bar{\Delta}(\om,\om_0)]^2\,,
&B[g_\tau] &= \sum_{\tau_1,\tau_2}
g_{\tau_1}g_{\tau_2}\operatorname{Cov}(\tau_1,\tau_2)\\
&\text{HLT:}
&A[g_\tau] &= \int_{\om_{\min}}^{\infty}d\om 
    |\Bar{\Delta}-\Delta_\sigma|^2\,,
&B[g_\tau] &= \sum_{\tau_1,\tau_2}
g_{\tau_1}g_{\tau_2}\operatorname{Cov}(\tau_1,\tau_2)\,
\end{align}
and $\Delta_\sigma$ is a gaussian with width $\sigma$ centred on
$\om_0$.  An important distinction between HLT and the other two is
that in HLT, the smearing width $\sigma$ is an input parameter, while
in the other two methods the smearing kernel $\smkernel{\om_0}$ and hence its
width is entirely determined by the data and the hyperparameter
$\lambda$.  The stability of the results with respect to variations in
$\lambda$ is an important consideration with these methods.

\begin{figure}[t]
\includegraphics[width=0.5\textwidth]{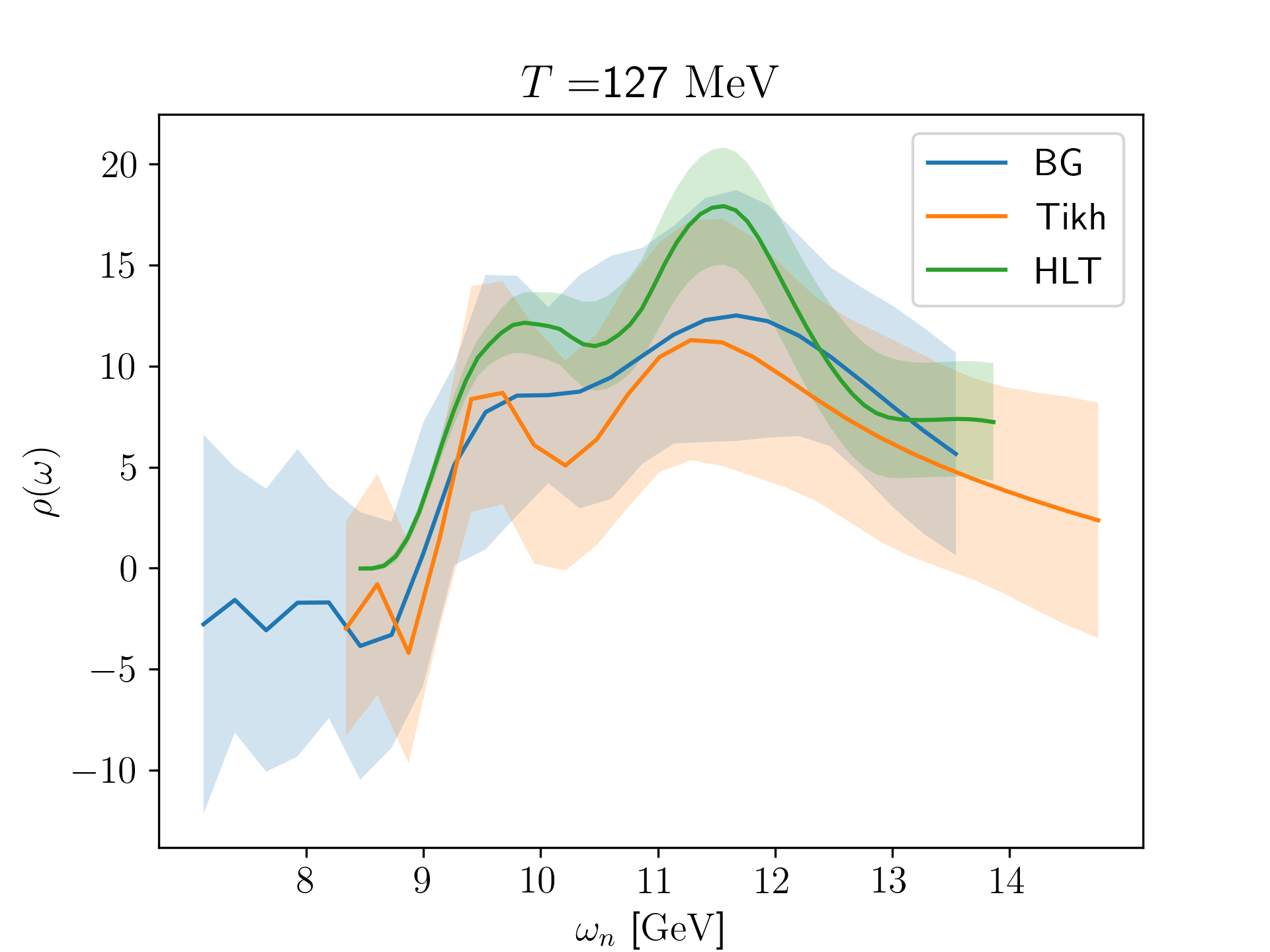}
\includegraphics[width=0.5\textwidth]{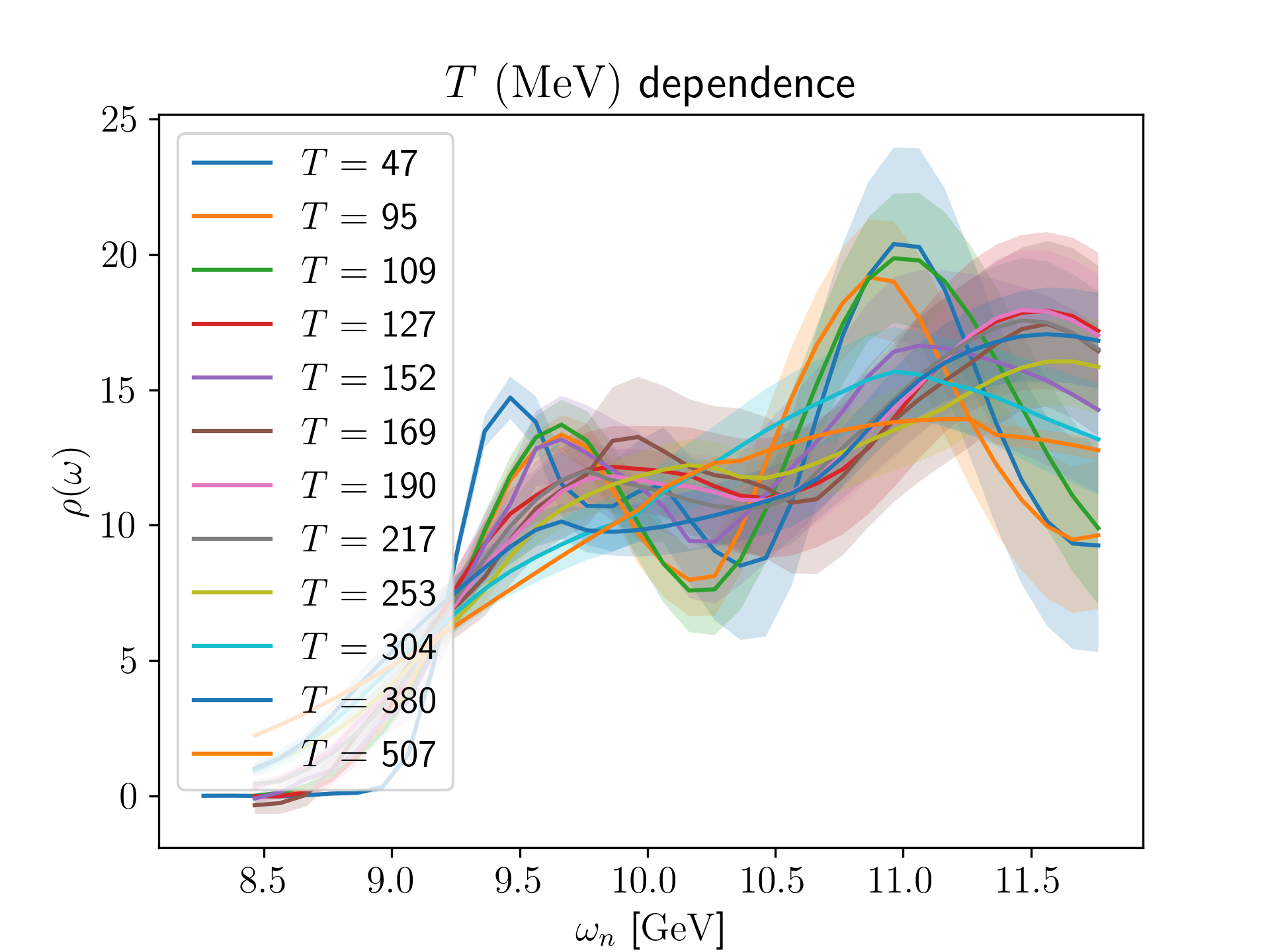}
\caption{Left: $\Upsilon$ spectral functions from Backus--Gilbert,
  Tikhonov and HLT methods at $T-127\,$MeV.  Right: HLT spectral
  functions at all temperatures.}
\label{fig:linear}
\end{figure}
Results for the $\Upsilon$ spectral functions from the three methods
are shown in the left panel of fig.~\ref{fig:linear}.  We see that the
results from all three methods are consistent, but that HLT yields a
considerably more well-determined spectral function than the other
two, as can be seen from the width of the respective error bands.

In the right panel of fig.~\ref{fig:linear} we show the spectral
functions from the HLT method at all temperatures.  We see a
progressive broadening and apparent positive mass shift as the
temperature is increased.  However, in light of the discussion in the
previous section, in the absence of an equivalent analysis on the
zero-temperature ensemble, we are not yet in a position to draw any
firm conclusion on this.
For more details about the methods and results, see \cite{Smecca:2025hfw}.

\section{Bayesian methods}
\label{sec:bayes}

Bayesian methods, including the maximum entropy method (MEM)
\cite{Asakawa:2000tr} and the BR method \cite{Burnier:2013nla},
regulate the inverse problem by introducing prior information,
such that the probability of the spectral function $\rho$ given the
data $D$ and the prior information $I$ is given by
\begin{equation}
  P(\rho|DI) = \frac{P(D|\rho I)P(\rho|I)}{P(D|I)}
  \propto e^{-L[D,\rho]+\alpha S[\rho]}\,,
\end{equation}
where $L$ is the standard likelihood ($\chi^2$) and $S[\rho]$
parametrises the prior probability $P(\rho|I)$.  $S$ is written in terms of a
\emph{default model} $m(\om)$ which is the most likely spectral
function in the absence of data.  The main difference between MEM and
BR lies in the form for $S[\rho]$,
\begin{align}
%  &\text{MEM:} &
  S_{\text{MEM}}[\rho] &= \int d\om\rho(\om)\ln\frac{\rho(\om)}{m(\om)}\,,
%  &\text{BR:}
 & S_{\text{BR}}[\rho] &=
  \int d\om \Big(1-\frac{\rho(\om)}{m(\om)} +
  \ln\frac{\rho(\om)}{m(\om)}\Big)\,.
\end{align}
The two methods also differ in that in the usual (Bryan)
implementation of MEM, $\rho(\om)$ is restricted to the singular
subspace of the kernel ($e^{-\om\tau}$ in the case of NRQCD), while BR
constructs the solution from the full $N_\om$-dimensional space.

\begin{figure}[t]
\includegraphics[width=0.5\textwidth]{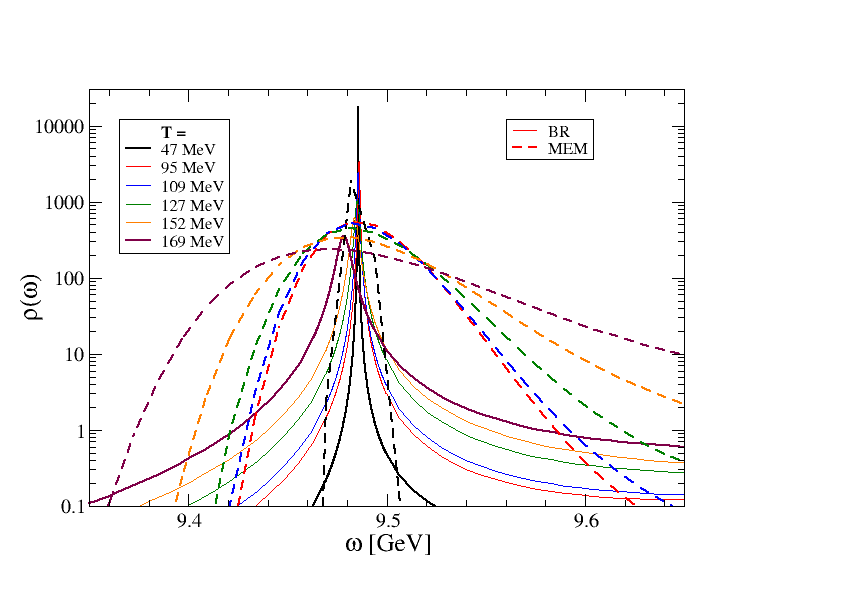}
\includegraphics[width=0.5\textwidth]{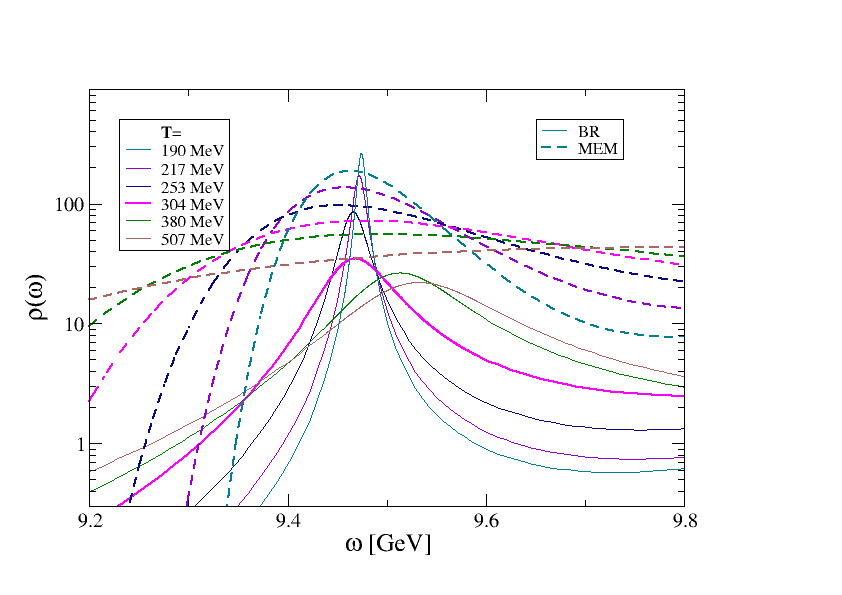}
\caption{$\Upsilon$ spectral functions from MEM (dashed lines) and BR (solid
  lines) at low (left) and high (right) temperature.  Note the
  difference in the scales in the two plots.}
\label{fig:MEM-BR-compare}
\end{figure}
In fig.~\ref{fig:MEM-BR-compare} we compare the results from MEM and
BR for the $\Upsilon$ spectral function in the vicinity of the
$\Upsilon$(1S) peak.  With both methods we see that the peak shifts to
the left and broadens as the temperature increases.  However, the
quantitative differences are significant: MEM consistently gives a
larger width than BR, and the shift in the peak position also tends to
be larger in the MEM results.

\begin{figure}[th]
\includegraphics[width=0.5\textwidth]{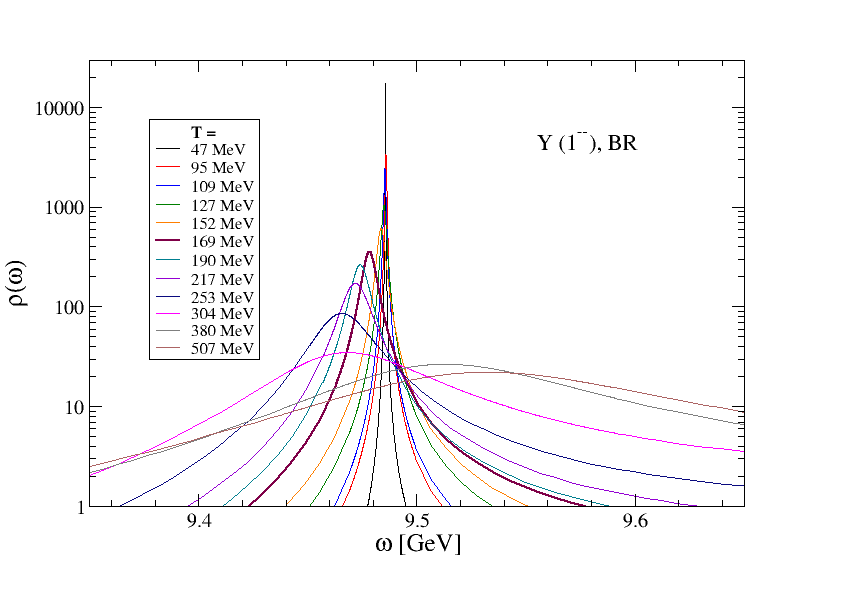}
\includegraphics[width=0.5\textwidth]{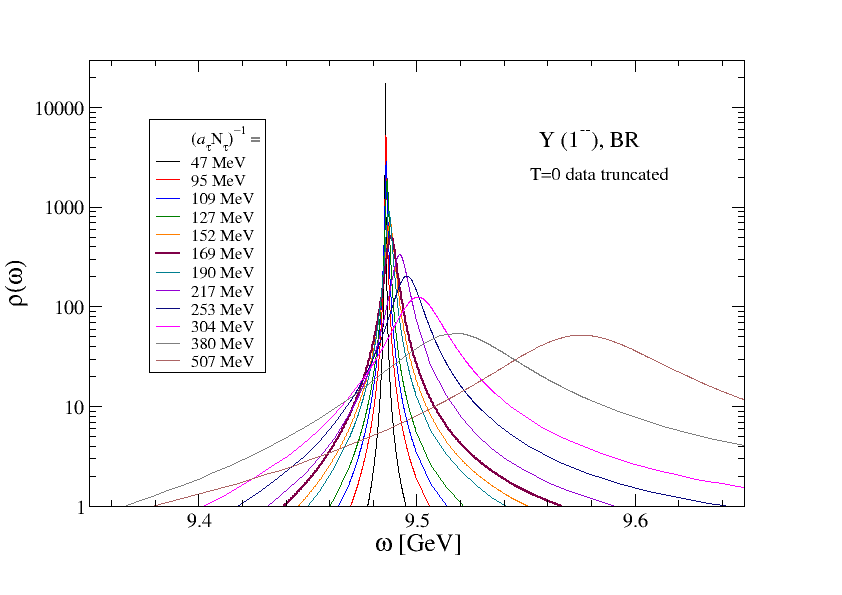} \\
\includegraphics[width=0.5\textwidth]{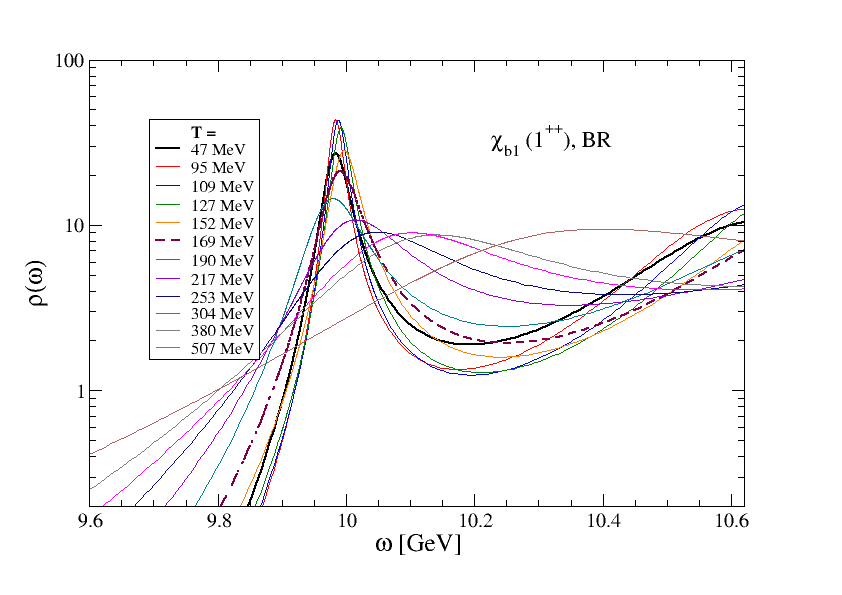}
\includegraphics[width=0.5\textwidth]{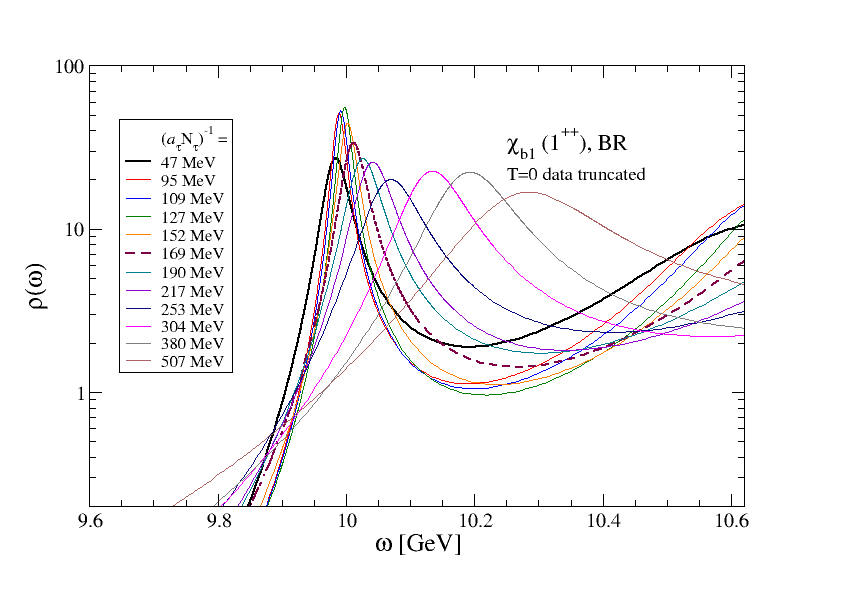}
\caption{Upsilon (top) and $\chi_{b1}$ spectral functions from the BR
  method.  The left hand plots show results from the thermal
  ensembles, while the right hand plots show results from the
  zero-temperature ensemble with the same temporal range.}
\label{fig:BR-spectral}
\end{figure}
Fig.~\ref{fig:BR-spectral} shows spectral functions from the BR
method for the $\Upsilon$ and $\chi_{b1}$ at all temperatures.  The
left hand plots show the results from the thermal ensembles, while the
right hand plots show results from the equivalent analysis on the
zero-temperature ensemble.  We see that the zero-temperature control
always gives a positive shift in the peak position and a broadening.
It is only when the thermal results differ from the control that a
real thermal effect can be concluded.  In this case we can infer
thermal effects in both channels; in particular, for the $\chi_{b1}$,
comparing the thermal and control results we can infer a negative mass
shift as well as a thermal broadening, while the thermal results on
their own would suggest a positive mass shift.

\section{Discussion}
\label{sec:comparison}

\begin{figure}[th]
\centering
\includegraphics[width=0.48\textwidth]{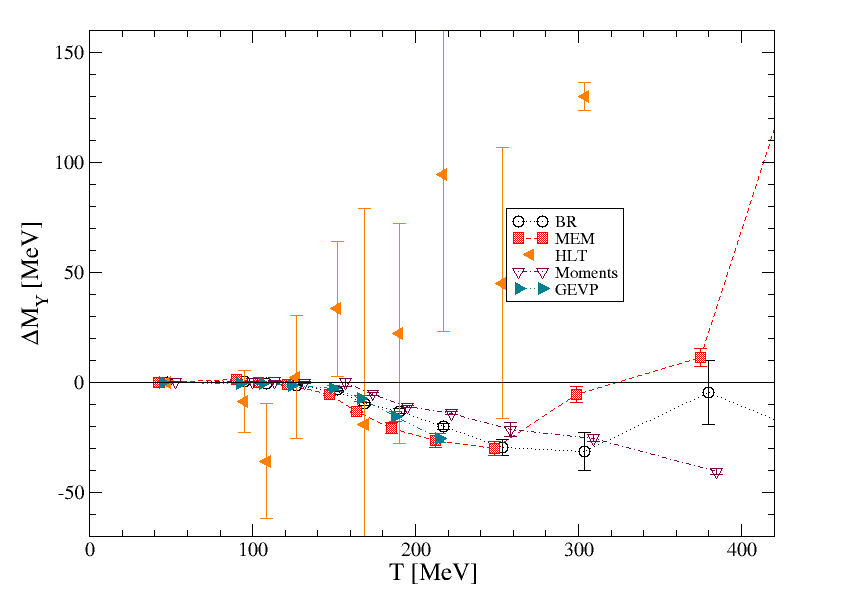}    
\includegraphics[width=0.48\textwidth]{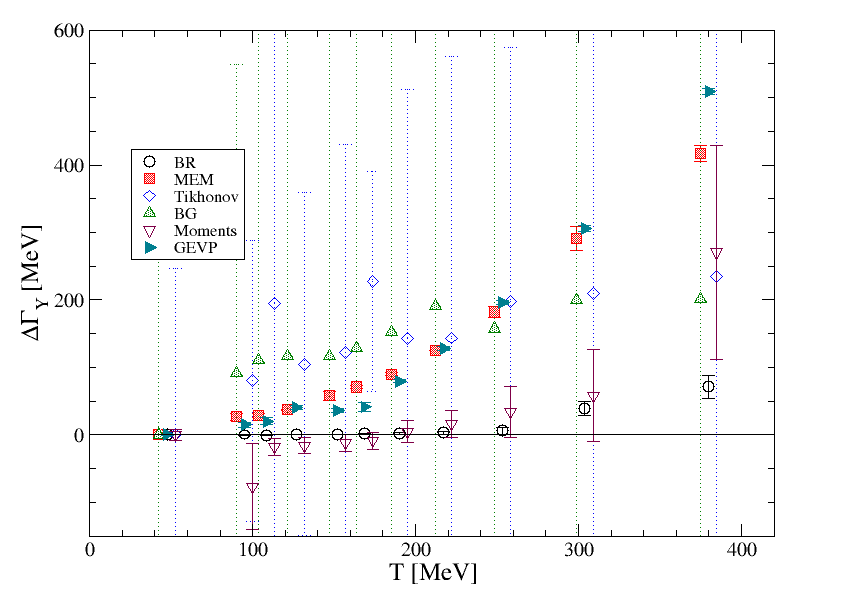}    \\
\includegraphics[width=0.48\textwidth]{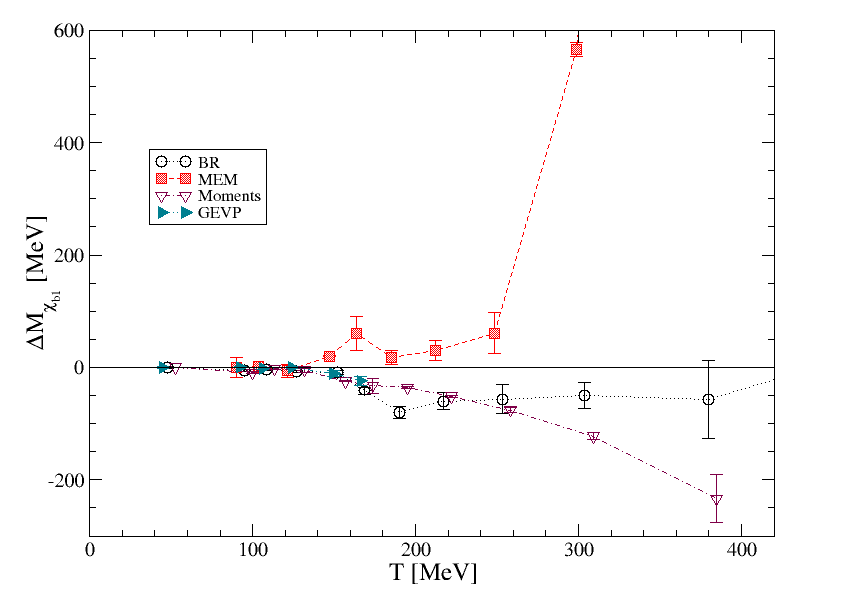}    
\includegraphics[width=0.48\textwidth]{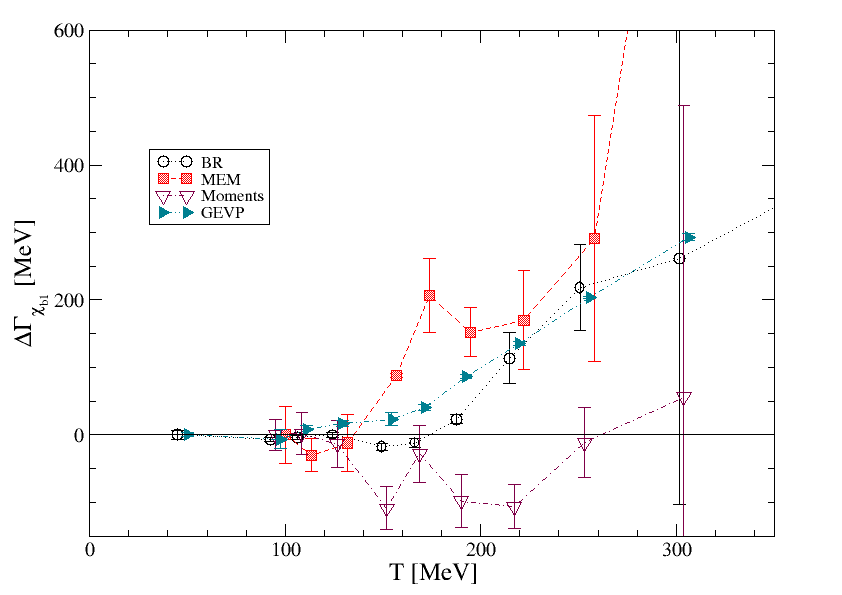}
\caption{Mass shift (left) and thermal width (right) for the
  $\Upsilon$(1S) (top) and $\chi_{b1}$(1P) (bottom), as functions of
  temperature $T$, for the different
methods presented in these proceedings.}
\label{fig:compare}
\end{figure}
Fig.~\ref{fig:compare} shows the results for the thermal mass shift
$\Delta M$ and thermal width $\Delta\Gamma$ of the $\Upsilon$ and
$\chi_{b1}$ for all the methods discussed here.  In addition, we show
results from  the use of a
generalised eigenvalue problem (GEVP) \cite{Bignell:2025bga}, where
masses are determined from standard exponential fits to the resulting
optimised correlators, while the widths are determined by applying the
time-derivative moments methods to the same correlators.

The mass shifts and thermal widths in fig.~\ref{fig:compare}
are obtained by subtracting the zero-temperature mass $M(T_0)$ and width
$\Gamma(T_0)$,
\begin{equation}
  \Delta M(T) = M(T)-M(T_0)\,,\qquad
  \Delta\Gamma(T) = \Gamma(T)-\Gamma(T_0)\,.
\end{equation}
For the moments and BR methods, $M(T_0)$ and $\Gamma(T_0)$ are
determined from the truncated zero-temperature correlators (control
data) as described in section~\ref{sec:moments}.  This prescription
attempts to correct for the main finite-$N_\tau$ effects, but it
should be noted that it may be complicated by the presence of excited
states at low $T$ which are not resolved in an analysis of the
truncated data and may not be present in the thermal correlators.
Further analysis will be necessary to disentange any such effects.

For the other methods (MEM, linear methods, GEVP) the ordinary
analysis at the lowest temperature has been used to determine
$M(T_0)$.  We expect the masses obtained from the GEVP method to be
relatively insensitive to finite-$N_\tau$ artefacts, while the GEVP
widths come from the moments method applied to the GEVP projected
correlators and so may have similar artefacts. We expect the MEM and
the linear methods to have a reduced $\Delta M$ and $\Delta\Gamma$
when the full control analysis is carried out.  In all cases,
uncertainties include both statistical and systematic uncertainties.

The first thing to note is that the linear methods have much larger
uncertainties than the others.  This is not surprising, as the width
of the smearing kernel in these methods is in most cases larger than
the physical width of the states under consideration.  The Tikhonov
and BG results for the $\Upsilon$ mass are too noisy to be shown.  In
HLT the width is an input parameter and can therefore not be used as
an estimator (or upper bound) for the physical width.

With these provisos, we see an encouraging qualitative and, in some
cases, even quantitative agreement between the different methods.
Notably, there is agreement between BR, MEM, moments and GEVP about a
small negative mass shift of up to 40\,MeV for the $\Upsilon$(1S), for
150\,MeV$\lesssim T\lesssim$250\,MeV.  A similar mass shift may be
seen in the $\chi_{b1}$(1P), using BR, moments and GEVP.  The current
MEM analysis does not show this shift, but the effect may be obscured
by the finite $N_\tau$, as observed with the BR method in
fig.~\ref{fig:BR-spectral}.

Concerning the width, there is rough agreement (within large
uncertainties) for the $\chi_{b1}$, but for the $\Upsilon$ there is a
discrepancy, with BR and moments giving a much smaller width
($\Delta\Gamma<100\,$MeV at $T=380\,$MeV) than the other methods.
Although the full zero-temperature control analysis may change this,
the GEVP method should be less sensitive to this, so this discrepancy
needs to be studied further.

\section{Summary and outlook}
\label{sec:summary}

We have studied the mass and width of ground state S- and P-wave
bottomonium ($\Upsilon$ and $\chi_{b1}$) with a range of different
methods.  We find agreement between correlator and Bayesian methods on
a negative $\Upsilon$ mass shift of up to 20--40\,MeV at
$T\sim250$\,MeV, as well as qualitative agreement on the mass and
width of $\chi_{b1}$.  There is a discrepancy for the width of
$\Upsilon$, which requires further investigation.  Linear methods
(Tikhonov, Backus--Gilbert, HLT) have intrinsically larger
uncertainties when applied to this particular problem, but their
results are consistent with those of the other methods.

Our results suggest that it is feasible to determine spectral features
at high temperature by comparing results using different methods.  A
crucial part of understanding the systematics is the zero-temperature
control analysis (see also \cite{Kelly:2018hsi,Kim:2018yhk}).
Applying this to all methods is currently in progress.  The procedure
outlined here may also be extended to study excited states
\cite{Bignell:2025bga}, or to transport properties, where the balance
of strengths and limitations of the different methods may be
different.

\section*{Data and software}

The gauge ensembles were produced using {\tt openqcd-fastsum}
\cite{openqcd-fastsum}; information about the ensembles can be found
at \cite{fastsum:gen2l}.  The NRQCD correlators were produced using
the \textsc{FASTNRQCD} package \cite{fastnrqcd}.  We anticipate making
the full dataset and scripts for the results presented here available
after a future publication.

\section*{Acknowledgments}

We acknowledge EuroHPC Joint Undertaking for awarding the project
EHPC-EXT-2023E01-010 access to LUMI-C, Finland. This work used the
DiRAC Data Intensive service (DIaL2 and DIaL) at the University of
Leicester, managed by the University of Leicester Research Computing
Service on behalf of the STFC DiRAC HPC Facility
(www.dirac.ac.uk). The DiRAC service at Leicester was funded by BEIS,
UKRI and STFC capital funding and STFC operations grants. This work
used the DiRAC Extreme Scaling service (Tesseract) at the University
of Edinburgh, managed by the Edinburgh Parallel Computing Centre on
behalf of the STFC DiRAC HPC Facility (www.dirac.ac.uk).  The DiRAC
service at Edinburgh was funded by BEIS, UKRI and STFC capital funding
and STFC operations grants. DiRAC is part of the UKRI Digital Research
Infrastructure. This work was performed using the PRACE Marconi-KNL
resources hosted by CINECA, Italy. We acknowledge the support of the
Supercomputing Wales project, which is part-funded by the European
Regional Development Fund (ERDF) via Welsh Government, and the
University of Southern Denmark for use of computing facilities. This work is
supported by STFC grant ST/X000648/1 and The Royal Society Newton
International Fellowship. RB acknowledges support from a Science
Foundation Ireland Frontiers for the Future Project award with grant
number SFI-21/FFP-P/10186. RHD acknowledges support from Taighde
Éireann – Research Ireland under Grant number GOIPG/2024/3507.

\bibliography{lattice,hot,jis}

\end{document}